\documentclass[11pt,english]{revtex4}
\usepackage{lmodern}

\usepackage[T1]{fontenc}
\usepackage[latin9]{inputenc}
\usepackage{amsmath}
\usepackage{amssymb}
\usepackage{esint}

\makeatletter
\@ifundefined{textcolor}{}
{%
 \definecolor{BLACK}{gray}{0}
 \definecolor{WHITE}{gray}{1}
 \definecolor{RED}{rgb}{1,0,0}
 \definecolor{GREEN}{rgb}{0,1,0}
 \definecolor{BLUE}{rgb}{0,0,1}
 \definecolor{CYAN}{cmyk}{1,0,0,0}
 \definecolor{MAGENTA}{cmyk}{0,1,0,0}
 \definecolor{YELLOW}{cmyk}{0,0,1,0}
 }

\usepackage{latexsym}\usepackage{bm}

\makeatother

\makeatother

\usepackage{babel}

\makeatother

\usepackage{babel}

\begin{document}

\title{Born-Infeld-Ho\v{r}ava gravity}

\author{\.{I}brahim Güllü }

\email{e075555@metu.edu.tr}

\affiliation{Department of Physics,\\
 Middle East Technical University, 06531, Ankara, Turkey}

\author{Tahsin Ça\u{g}r\i{} \c{S}i\c{s}man}

\email{sisman@metu.edu.tr}

\affiliation{Department of Physics,\\
 Middle East Technical University, 06531, Ankara, Turkey}

\author{Bayram Tekin}

\email{btekin@metu.edu.tr}

\affiliation{Department of Physics,\\
 Middle East Technical University, 06531, Ankara, Turkey}

\date{\today}
\begin{abstract}
We define various Born-Infeld gravity theories in $3+1$ dimensions
which reduce to Ho\v{r}ava's model at the quadratic level in small
curvature expansion. In their exact forms, our actions provide $z\rightarrow\infty$
extensions of Ho\v{r}ava's gravity, but when small curvature expansion
is used, they reproduce finite $z$ models, including some half-integer
ones. 
\end{abstract}
\maketitle

\section{Introduction}

Born-Infeld (BI) \cite{bi} type actions appear in physics in various
contexts; for example, the simplest one is the relativistic point
particle action $I=-m\int dt\,\sqrt{1-v^{2}}$. One pragmatic way
of looking at this action is that it restricts $v\leq1$. Similarly,
in electrodynamics to put an upper bound to the electric field, and
obtain a finite self-energy for the point charge, electrodynamics
can be extended to \begin{equation}
I=-b^{2}\int d^{4}x\,\sqrt{-\det\left(g_{\mu\nu}+\frac{1}{b}F_{\mu\nu}\right)},\label{eq:BI_original}\end{equation}
 where $b$ is a dimensionful parameter which sets the scale of the
maximum attainable electric field. It is easy to check that at the
quadratic order, after dropping a constant term, (\ref{eq:BI_original})
gives the pure Maxwell theory. It also has the desired properties
such as ghost freedom and causal propagation. In string theory, Nambu-Goto
action and D-brane actions are of the BI type. For a nice account
of BI theories see \cite{gibbons}. It is only natural to consider
determinantal actions in gravity theories. In fact, a decade before
the BI paper, Eddington \cite{eddington}, using the symmetric connection
(not the metric) as the independent field, extended general relativity
in the form $I\sim\int d^{4}x\,\sqrt{\det\left[R_{\left(\mu\nu\right)}\left(\Gamma\right)\right]}$
which has recently picked up interest \cite{banados}. Deser and Gibbons
\cite{gibbonsDeser}, using the metric as the independent field, pondered
upon viable BI-type gravity models in four dimensions. They considered
the action \[
I=\int d^{4}x\,\sqrt{-\det\left(ag_{\mu\nu}+bR_{\mu\nu}+cX_{\mu\nu}\right)},\]
 where $X_{\mu\nu}$ is a {}``fudge'' tensor which should be chosen
in such a way that ghosts and tachyons do not appear in the small
curvature expansion about the Minkowski or (anti)-de Sitter spaces.

BI-type consistent gravity actions take their most elegant form in
three dimensions. Recently \cite{gst}, it was shown that the action\begin{equation}
I_{BI}=-\frac{4m^{2}}{\kappa^{2}}\int d^{3}x\,\left[\sqrt{-\det\left(-\frac{1}{m^{2}}\mathcal{G}\right)}-\left(\frac{\Lambda}{2m^{2}}+1\right)\sqrt{-g}\right],\label{eq:BI-NMG}\end{equation}
 with $\mathcal{G}_{\mu\nu}\equiv R_{\mu\nu}-\frac{1}{2}g_{\mu\nu}R-m^{2}g_{\mu\nu}$,
at the quadratic level reduces to the new massive gravity (NMG)\cite{bht}
\[
I_{NMG}=\frac{1}{\kappa^{2}}\int d^{3}x\,\sqrt{-g}\left[-\left(R-2\Lambda\right)+\frac{1}{m^{2}}\left(R_{\mu\nu}^{2}-\frac{3}{8}R^{2}\right)\right],\]
 which is the only unitary \cite{bht,nakasone,deser,gullu}, super-renormalizable
\cite{stelle}, parity-invariant theory. Conforming to the spirit
of the BI actions, for constant curvature spaces, (\ref{eq:BI-NMG})
restricts the curvature to be $R_{\mu\nu}\ge-2m^{2}g_{\mu\nu}$. At
the cubic order, (\ref{eq:BI-NMG}) yields the action \begin{align}
I_{\text{NMG extended}} & =\frac{1}{\kappa^{2}}\int d^{3}x\,\sqrt{-g}\left[-\left(R-2\Lambda\right)+\frac{1}{m^{2}}\left(R_{\mu\nu}^{2}-\frac{3}{8}R^{2}\right)\right.\nonumber \\
 & \phantom{=\frac{1}{\kappa^{2}}\int d^{3}x\,\sqrt{-g}}\left.+\frac{2}{3m^{4}}\left(R^{\mu\nu}R_{\nu}^{\phantom{\nu}\alpha}R_{\alpha\mu}-\frac{9}{8}RR_{\mu\nu}^{2}+\frac{17}{64}R^{3}\right)+O\left(R^{4}\right)\right],\label{eq:NMG-ext}\end{align}
 which exactly matches the action obtained by Sinha \cite{sinha}
who used the AdS/CFT conjecture and the existence of a holographic
$c$ theorem to find cubic deformations to NMG. It is remarkable that
a BI-type gravitational action at the cubic order reproduces a three-dimensional
theory in the bulk which is fixed by conformal field theory on the
two-dimensional boundary.

Inspired by the success of the BI actions, in this work we will present
various extensions of Ho\v{r}ava's recent nonrelativistic gravity
theory \cite{horava} to all orders in the curvature. The layout of
the paper is as follows: In Sec. \ref{sec:BI-Type-Generating-Action},
we propose an extension of Ho\v{r}ava's gravity by defining a BI-type
potential generating action in three dimensions using the detailed-balance
principle. In Sec. \ref{sec:BI-Type-Potential}, without reference
to the detailed-balance principle, we give BI-type extensions of Ho\v{r}ava's
gravity in $3+1$ dimensions.

\section{Born-Infeld-Ho\v{r}ava Gravity: BI-Type Potential Generating Action\label{sec:BI-Type-Generating-Action}}

Since Ho\v{r}ava's gravity has already been described in many places
such as \cite{weinfurtner}, just to fix the notation, using (almost)
the form of the action given in \cite{lu}, we shall briefly recapitulate
the essential ingredients. (See \cite{HL-refs} for related works.)
One starts with the usual ADM \cite{ADM} decomposition of the four-dimensional
space, \[
ds^{2}=-N^{2}dt^{2}+g_{ij}\left(dx^{i}-N^{i}dt\right)\left(dx^{j}-N^{j}dt\right),\]
 where all the involved functions depend on $t$ and $x_{i}$. {[}Note
that we do not commit ourselves to the so-called projectable version
of Ho\v{r}ava's gravity for which $N$ depends on $t$ only.{]} One
then assumes different scaling dimensions for time and space: $t\rightarrow b^{z}t$,
$x^{i}\rightarrow bx^{i}$. It is expected that in the IR limit $z\rightarrow1$
and full diffeomorphism invariance is recovered. Once sacred Lorentz
invariance is let go, there is no limit to the number of models one
can define. Ho\v{r}ava introduced a guiding principle called the {}``detailed
balance'' to inherit {}``reasonable'' actions from three dimensions.
In short, his specific proposal leads to \[
I_{H}=\int dtd^{3}\boldsymbol{x}\,\sqrt{g}N\left(\mathcal{L}_{K}+\mathcal{L}_{V}\right),\]
 where $\mathcal{L}_{K}$ and $\mathcal{L}_{V}$ are kinetic and potential
parts, respectively. Not to introduce more than two time derivatives
and get hit by the Ostragradski ghosts, the kinetic part is defined
as\[
\mathcal{L}_{K}=\frac{2}{\kappa^{2}}\left(K_{ij}K^{ij}-\lambda K^{2}\right),\]
 where $K_{ij}=\frac{1}{2N}\left(\dot{g}_{ij}-\nabla_{i}N_{j}-\nabla_{j}N_{i}\right)$
is the extrinsic curvature and $\lambda$ is a dimensionless coupling
constant which hopefully flows to $1$ in the IR limit, so that one
recovers the kinetic part of the standard Einstein-Hilbert action.
The kinetic part of the action is pretty robust, but as for the potential
part, one has a great deal of freedom. Specifically, defining $C^{ij}=\epsilon^{ikl}\nabla_{k}\left(R_{\phantom{j}l}^{j}-\frac{1}{4}R\delta_{l}^{j}\right)$
to be the Cotton tensor, the $z=3$ Ho\v{r}ava theory has the potential:\[
\mathcal{L}_{V}=\frac{\kappa^{2}\mu^{2}\Lambda_{W}\left(R-3\Lambda_{W}\right)}{8\left(1-3\lambda\right)}+\frac{\kappa^{2}\mu^{2}\left(1-4\lambda\right)}{32\left(1-3\lambda\right)}R^{2}-\frac{\kappa^{2}}{2w^{4}}\left(C_{ij}-\frac{\mu w^{2}}{2}R_{ij}\right)\left(C^{ij}-\frac{\mu w^{2}}{2}R^{ij}\right),\]
 where \emph{à la }the detailed-balance principle, $\mathcal{L}_{V}$
comes from the three-dimensional Einstein-Hilbert and the topologically
massive gravity (TMG) actions. More concretely, this principle works
in the following way, $\mathcal{L}_{V}=\frac{\kappa^{2}}{8}E^{ij}G_{ijkl}E^{kl}$,
where $E^{ij}=\frac{1}{\sqrt{g}}\frac{\delta W_{3}}{\delta g_{ij}}$.
Here, $G_{ijkl}$ is the deformed De Witt metric \[
G_{ijkl}=\frac{1}{2}\left(g_{ik}g_{jl}+g_{il}g_{jk}\right)+\frac{\lambda}{1-3\lambda}g_{ik}g_{jl}\]
 and $W_{3}$ is a three-dimensional Euclidean action. {[}As a side
remark, note that in a slightly different context, using the method
of steepest descent topologically massive gravity was used to obtain
the $C_{ij}C^{ij}$ action to define another nonrelativistic theory
which is the Cotton flow theory $\partial_{t}g_{ij}=C_{ij}$ \cite{kisisel}.
The relevance of Cotton and related flows to Ho\v{r}ava's gravity
has been recently studied in \cite{bakas}.{]}

In \cite{cai}, $z=4$ Ho\v{r}ava's theory was defined using the NMG
as the potential generating action via the detailed-balance principle.
As we noted in the Introduction, NMG itself has a consistent BI-type
extension. Therefore, our first proposal is to add to TMG the Euclidean
BI action to get the potential generating action:\begin{align}
W_{3} & =-4\mu^{2}L\int d^{3}x\,\left[\sqrt{\det\left(\frac{1}{\mu L}\mathcal{G}\right)}+\left(\frac{\Lambda_{W}}{2\mu L}-1\right)\sqrt{g}\right]\nonumber \\
 & \phantom{=}+\frac{1}{w^{2}}\int d^{3}x\,\sqrt{g}\epsilon^{ijk}\Gamma_{il}^{m}\left(\partial_{j}\Gamma_{km}^{l}+\frac{2}{3}\Gamma_{jn}^{l}\Gamma_{km}^{n}\right),\label{eq:Pot_gen_act}\end{align}
 where $\mathcal{G}_{ij}\equiv R_{ij}-\frac{1}{2}g_{ij}R+\mu Lg_{ij}$.
This action can be used to obtain the potential at any desired order.
In the quadratic order, the result of \cite{cai} follows, at the
cubic order one should find the equations coming from (\ref{eq:NMG-ext})
add the Cotton tensor to find $E^{ij}$, and using the De Witt metric
described as above take the square of $E_{ij}$ to get the potential.
If one wants to deal with the exact (that is the $z\rightarrow\infty$)
theory, one can use\[
\det A=\frac{1}{6}\left[\left(\text{Tr}A\right)^{3}-3\text{Tr}A\text{Tr}\left(A^{2}\right)+2\text{Tr}\left(A^{3}\right)\right],\]
 to get\begin{align*}
\sqrt{\det\left(\frac{1}{m^{2}}\mathcal{G}\right)} & =\sqrt{g}\left(1-\frac{1}{2m^{2}}R^{ij}\left[g_{ij}+\frac{1}{m^{2}}\left(R_{ij}-\frac{1}{2}g_{ij}R\right)\right.\right.\\
 & \phantom{=\sqrt{g}\left\{ 1-\frac{1}{2m^{2}}R^{ij}\right.}\left.\left.-\frac{2}{3m^{4}}\left(R_{ik}R_{\phantom{\alpha}j}^{k}-\frac{3}{4}RR_{ij}+\frac{1}{8}g_{ij}R^{2}\right)\right]\right)^{1/2},\end{align*}
 or if one wants to work at a finite $z$, for example, such as $z=8$
one can do a small curvature expansion up to $O\left(A^{5}\right)$,
\begin{align*}
\left[\det\left(1+A\right)\right]^{1/2} & =1+\frac{1}{2}\text{Tr}A-\frac{1}{4}\text{Tr}\left(A^{2}\right)+\frac{1}{8}\left(\text{Tr}A\right)^{2}+\frac{1}{6}\text{Tr}\left(A^{3}\right)-\frac{1}{8}\text{Tr}A\text{Tr}\left(A^{2}\right)+\frac{1}{48}\left(\text{Tr}A\right)^{3}\\
 & \phantom{=}-\frac{1}{8}\text{Tr}\left(A^{4}\right)+\frac{1}{32}\left[\text{Tr}\left(A^{2}\right)\right]^{2}+\frac{1}{12}\text{Tr}A\text{Tr}\left(A^{3}\right)-\frac{1}{32}\left(\text{Tr}A\right)^{2}\text{Tr}\left(A^{2}\right)+\frac{1}{384}\left(\text{Tr}A\right)^{4}.\end{align*}
 Clearly, this procedure can be extended to any desired order.

\section{Born-Infeld-Ho\v{r}ava Gravity: BI-Type Potential\label{sec:BI-Type-Potential}}

In the above section, we proposed that the potential generating three-dimensional
action for Ho\v{r}ava's gravity can be taken to be the gravitational
BI action together with the TMG action. This procedure leads to a
manageable deformation of Ho\v{r}ava's gravity. In this section, we
will propose a more radical extension of Ho\v{r}ava's gravity again
in the form of a BI action which will not require a reference to the
detailed-balance principle. First, observe that for $\lambda=1$ the
potential part of Ho\v{r}ava's theory reduces to the Euclidean NMG
in addition to Cotton parts:\begin{align}
\mathcal{L}_{V} & =-\frac{\kappa^{2}\mu^{2}\Lambda_{W}}{16}\left[\left(R-3\Lambda_{W}\right)+\frac{2}{\Lambda_{W}}\left(R_{ij}^{2}-\frac{3}{8}R^{2}\right)\right]-\frac{\kappa^{2}}{2w^{4}}C_{ij}^{2}+\frac{\kappa^{2}\mu}{2w^{2}}C^{ij}R_{ij}.\label{eq:Horava_pot_l_is_1}\end{align}
 Note that the appearance of NMG in the IR limit of $z=3$ Ho\v{r}ava's
gravity should not be confused with the use of NMG as a potential
generating action for the $z=4$ theory discussed above. Observation
of (\ref{eq:Horava_pot_l_is_1}) led us to consider the following
BI action: \begin{align}
I_{BI} & =\frac{2}{\kappa^{2}}\int dtd^{3}\boldsymbol{x}\,\sqrt{g}N\left(K_{ij}K^{ij}-\lambda K^{2}\right)\nonumber \\
 & \phantom{=}+\frac{1}{b}\int dtd^{3}x\, N\left\{ \sqrt{\det\left[g_{ij}+a\tilde{R}_{ij}+dg_{ij}R+eC_{ij}\right]}+\frac{1}{2}\sqrt{g}\right\} ,\label{eq:BI_ext_HI_1}\end{align}
 where $\tilde{R}_{ij}\equiv R_{ij}-\frac{1}{3}g_{ij}R$. With the
coefficients\[
a=\pm\frac{\sqrt{6\lambda-2}}{\Lambda_{W}},\quad\lambda>\frac{1}{3},\qquad b=\frac{2a^{2}}{\kappa^{2}\mu^{2}},\qquad d=-\frac{1}{3\Lambda_{W}},\qquad e=-\frac{2a}{\mu w^{2}},\]
 Ho\v{r}ava's gravity is reproduced in the small curvature expansion
at the quadratic level. What is of course remarkable about (\ref{eq:BI_ext_HI_1})
is that by just considering the metric, the Ricci tensor and scalar
and the Cotton tensor and not any other higher derivative tensors,
one can reproduce and extend Ho\v{r}ava's gravity to any order. For
example, at $O\left(R^{3}\right)$ Ho\v{r}ava's action will be augmented
with \begin{align}
b\mathcal{L}_{O\left(R^{3}\right)} & =\frac{1}{6}\left(a^{3}R^{ij}R_{jk}R_{\phantom{k}i}^{k}+3a^{2}eR^{ij}R_{jk}C_{\phantom{k}i}^{k}+3ae^{2}R^{ij}C_{jk}C_{\phantom{k}i}^{k}+e^{3}C^{ij}C_{jk}C_{\phantom{k}i}^{k}\right)\nonumber \\
 & \phantom{=}-\frac{1}{6}\left(a-\frac{3}{4}d\right)R\left(a^{2}R_{ij}^{2}+2aeR^{ij}C_{ij}+e^{2}C_{ij}^{2}\right)+\left(\frac{a^{3}}{27}-\frac{1}{24}a^{2}d-\frac{1}{16}d^{3}\right)R^{3},\label{eq:OR3_ext}\end{align}
 which defines a $z=4.5$ theory in the UV. Beyond this order, the
computation gets more cumbersome (see the Appendix).

We stress that the BI action (\ref{eq:BI_ext_HI_1}) is tailor made
to reproduce Ho\v{r}ava's gravity at the quadratic order. Therefore,
as long as one considers small curvature expansions at the desired
order, Ho\v{r}ava's theory merely receives corrections. For example,
the solutions found in \cite{lu,kiritsis} will be modified. On the
other hand, considering (\ref{eq:BI_ext_HI_1}) as the \emph{exact}
theory without any approximation, one can search for solutions. It
is easy to check that spherically symmetric \emph{static} solutions
are ruled out. To see this, leaning on \emph{symmetric} \emph{criticality},
which says that symmetric critical points are critical symmetric points
when compact symmetry group is integrated out \cite{palais,deserTekin},
one inserts the ansatz\[
ds^{2}=-N^{2}\left(r\right)dt^{2}+\frac{1}{f\left(r\right)}dr^{2}+r^{2}\left(d\theta^{2}+\sin^{2}\theta d\varphi^{2}\right),\]
 to the action (\ref{eq:BI_ext_HI_1}): The kinetic part vanishes
and variation with respect to $N\left(r\right)$ shows that there
cannot be a static solution. This result is surprising, but (\ref{eq:BI_ext_HI_1})
is supposed to define a quantum gravity action and there is no guarantee
that in quantum gravity there will be spherically symmetric \emph{static}
solutions. In fact, in Einstein's gravity the classical Schwarzschild
solution fails to be static even in the semiclassical approach \cite{hawking}:
It has Hawking radiation.

Finally, by allowing quadratic terms inside the determinant, one can
find some nonminimal BI extensions of Ho\v{r}ava's gravity. The most
general choice using only the Cotton and the Ricci tensors would be
\begin{multline*}
\det\left[g_{ij}+a\tilde{R}_{ij}+dg_{ij}R+eC_{ij}+f\left(RR_{ij}+lg_{ij}R^{2}\right)+n_{R}\left(R_{ik}R_{\phantom{k}j}^{k}+p_{R}g_{ij}R_{kl}^{2}\right)\right.\\
\left.+n_{RC}\left(R_{ik}C_{\phantom{k}j}^{k}+p_{RC}g_{ij}R_{kl}C^{kl}\right)+n_{C}\left(C_{ik}C_{\phantom{k}j}^{k}+p_{C}g_{ij}C_{kl}^{2}\right)\right].\end{multline*}
 The requirement that Ho\v{r}ava's gravity be reproduced at the quadratic
level is highly restrictive, and the explicit computation shows that
not all the terms are allowed. In fact, one is left to choose either
the $RR_{ij}$ or the $g_{ij}R^{2}$ term. Therefore, another extension
of Ho\v{r}ava's gravity is \begin{align}
I_{\text{BI nonminimal}} & =\frac{2}{\kappa^{2}}\int dtd^{3}\boldsymbol{x}\,\sqrt{g}N\left(K_{ij}K^{ij}-\lambda K^{2}\right)\nonumber \\
 & \phantom{=}+\frac{1}{b}\int dtd^{3}x\, N\sqrt{\det\left[g_{ij}+a\tilde{R}_{ij}+dg_{ij}R+eC_{ij}+fR_{ij}R\right]},\label{eq:Horava_non-minimal}\end{align}
 where\[
a=\pm\frac{2}{\Lambda_{W}}\sqrt{\frac{3\lambda-1}{3}},\qquad b=\frac{2a^{2}}{\kappa^{2}\mu^{2}},\qquad d=-\frac{2}{9\Lambda_{W}},\qquad e=-\frac{2a}{\mu w^{2}},\qquad f=\frac{1}{54\Lambda_{W}^{2}}.\]
 Observe that, as opposed to (\ref{eq:BI_ext_HI_1}), in (\ref{eq:Horava_non-minimal})
the cosmological constant term comes with the correct factors from
the determinant and one does not need to add a $\sqrt{g}$. In principle,
this action allows spherically symmetric static solutions. Again small
curvature expansion will give deformations of Ho\v{r}ava's theory
at any order.

\section{Conclusions}

We proposed, without introducing new parameters, three BI-type extensions,
(\ref{eq:Pot_gen_act}), (\ref{eq:BI_ext_HI_1}), (\ref{eq:Horava_non-minimal}),
of Ho\v{r}ava's gravity. All these extensions can be used to generate
finite $z$ theories, or taken in their exact form they define in
a compact way $z\rightarrow\infty$ theories in the UV regimes. Our
first proposal (\ref{eq:Pot_gen_act}) uses the detailed-balance principle,
and the potential is generated from a three-dimensional BI action
that generalizes the NMG and its cubic deformation obtained from AdS/CFT.
The other two proposals do not use the detailed-balance principle:
the potential is given in terms of a BI-type action. In the literature,
both the original Ho\v{r}ava's gravity and its extensions \cite{sotiriou,blas}
have been questioned regarding their consistency and strong coupling
problems \cite{charmousis,papazoglou,kimpton}. The models we have
proposed here need to be studied along these lines. Classical solutions
of the BI-type theories have usually no or less severe singularities.
It would be interesting to see if our actions have nonsingular solutions.

\section*{\label{ackno} Acknowledgments}

T.Ç.\c{S}. is supported by a T{Ü}B\.{I}TAK Ph.D. Scholarship. B.T.
is partially supported by the T{Ü}B\.{I}TAK Kariyer Grant No. 104T177.

\section*{Appendix: $O\left(R^{4}\right)$ extension of BI-Ho\v{r}ava Gravity}

In addition to (\ref{eq:OR3_ext}), the following terms will be added
to get the $z=6$ theory at $O\left(R^{4}\right)$:

\begin{align*}
b\mathcal{L}_{O\left(R^{4}\right)} & =-\frac{1}{8}\left(a^{4}R^{ij}R_{jk}R^{kl}R_{li}+4a^{3}eR^{ij}R_{jk}R^{kl}C_{li}+6a^{2}e^{2}R^{ij}R_{jk}C^{kl}C_{li}\right.\\
 & \phantom{=-\frac{1}{8}}\left.+4ae^{3}R^{ij}C_{jk}C^{kl}C_{li}+e^{4}C^{ij}C_{jk}C^{kl}C_{li}\right)\\
 & \phantom{=}+\frac{1}{12}\left(2a-3d\right)R\left(a^{3}R^{ij}R_{jk}R_{\phantom{k}i}^{k}+3a^{2}eR^{ij}R_{jk}C_{\phantom{k}i}^{k}+3ae^{2}R^{ij}C_{jk}C_{\phantom{k}i}^{k}+e^{3}C^{ij}C_{jk}C_{\phantom{k}i}^{k}\right)\\
 & \phantom{=}-\frac{1}{96}\left(9d^{2}-24ad+10a^{2}\right)R^{2}\left(a^{2}R_{ij}^{2}+2aeR^{ij}C_{ij}+e^{2}C_{ij}^{2}\right)\\
 & \phantom{=}+\frac{1}{32}\left[a^{4}\left(R_{ij}^{2}\right)^{2}+4a^{3}eR_{ij}^{2}R^{kl}C_{kl}+2a^{2}e^{2}R_{ij}^{2}C_{kl}^{2}\right.\\
 & \phantom{=+\frac{1}{32}}\left.+4a^{2}e^{2}\left(R^{ij}C_{ij}\right)^{2}+4ae^{3}R^{ij}C_{ij}C_{kl}^{2}+e^{4}\left(C_{ij}^{2}\right)^{2}\right]\\
 & \phantom{=}+\frac{1}{32}\left(\frac{5a^{4}}{9}-\frac{16a^{3}d}{9}+a^{2}d^{2}+\frac{3d^{4}}{4}\right)R^{4}.\end{align*}

\end{document}